\documentclass[12pt,preprint]{aastex}
\usepackage{amsmath}

\shorttitle{Subarcsecond Structure of B335}
\shortauthors{Harvey et al.}

\begin{document}

\title{Disk Properties and Density Structure of the Star-Forming Dense
Core B335
\footnote{Based on observations carried out with the IRAM Plateau de Bure 
Interferometer.  IRAM is supported by INSU/CNRS (France), MPG (Germany) 
and IGN (Spain).}
}

\author{Daniel W.A.\ Harvey, David J.\ Wilner, Philip C.\ Myers}

\email{dharvey, dwilner, pmyers@cfa.harvard.edu}

\affil{Harvard-Smithsonian Center for Astrophysics, 60 Garden Street,
Cambridge, MA 02138}

\and

\author{Mario Tafalla}

\email{m.tafalla@oan.es}

\affil{Observatorio Astron\'{o}mico Nacional, Alfonso XII 3, E-28014
Madrid, Spain}

\begin{abstract}
We present subarcsecond resolution dust continuum observations 
of the protostellar collapse candidate B335 made with the IRAM 
Plateau de Bure Interferometer at wavelengths of 1.2 and 3.0~mm.
These observations probe to $<100$~AU size scales and reveal a 
compact source component that we identify with a circumstellar disk.  
We analyze these data in concert with previous lower resolution 
interferometer observations and find a best fit density structure 
for B335 that consists of a power law envelope with index 
$p=1.55 \pm 0.04$ ($r \lesssim 5000$~AU) together with a disk
($r < 100$~AU) of flux $F_{1.2\mathrm{mm}}=21 \pm 2$~mJy. 
We estimate a systematic uncertainty in the power law index 
$\delta p \lesssim 0.15$, where the largest error comes from 
the assumed form of the dust temperature falloff with radius.
This determination of the inner density structure of B335 has 
a precision unique amongst protostellar cores, and it is consistent 
with the $r^{-1.5}$ profile of gravitational free-fall, 
in accord with basic expectations for the formation of a star.
The flux (and implied mass) of the compact component
in B335 is typical of the disks around T~Tauri stars.
\end{abstract}

\keywords{ISM: globules --- ISM: individual(B335) 
--- radio continuum: ISM
--- stars: formation 
---  accretion, accretion disks}

\section{Introduction}
The dense core in the B335 dark globule is generally recognized as
the best protostellar collapse candidate. This dense core is nearby
(250~pc, Tomita, Saito \& Ohtani 1979), isolated, and nearly spherical.
It contains a deeply embedded low luminosity young stellar object
($3$~L${}_{\odot}$) discovered at far-infrared wavelengths by
Keene et al.\ (1983) and detected by IRAS only at $\lambda \geq 60$~$\mu$m.

Detailed radiative transfer models based on the theory of
inside-out collapse (Shu 1977) provide very good fits to
spectral line profiles of the dense gas tracers CS and H$_2$CO observed
at $10''$ to $30''$ resolution (Zhou et al.\ 1993; Choi et al.\ 1995),
though recent studies have raised doubts about the inside-out collapse
interpretation. Observations by Wilner et al.\ (2000) of the CS(5-4) line 
at higher angular resolution, $\sim2\farcs5$ ($\sim$500~AU), show that the 
high velocity emission arises from the inner part of a bipolar outflow, 
not from the gravitational acceleration of gas close to the protostar. 
However, studies of starless cores show that some molecular species, in 
particular CS, are frozen out onto grains at densities characteristic of 
the inner regions of B335 (Caselli et al.\ 2002, Tafalla et al.\ 2002). 
If heating does not promptly desorb CS molecules from grain mantles, then 
this species likely makes a poor probe of the kinematics in the presumed 
infall zone. 

Since the density field is strongly coupled to the velocity field, 
observations of long wavelength dust emission provide an alternative means 
to infer the dynamical state of the dense core. The intensity of the 
dust emission provides an integral along the line-of-sight of the product 
of the density and dust temperature. By modeling the dust temperature and
specific mass opacity, the observed intensities can be used to constrain
the density distribution. This approach has been applied extensively
using large single-dish telescopes (e.g. Ward-Thompson, Motte \& Andre 1999, 
Motte \& Andre 2001, Visser, Richer \& Chandler 2001, Shirley et al.\ 2000). 
But the central regions of the nearest protostellar cores are generally
comparable in size to the beamwidths of single-dish telescopes and
remain poorly resolved. The density structure at smaller scales must be 
probed using interferometers. So far, detailed modeling of millimeter 
interferometer data in this context has been rare (Keene \& Masson 1990, 
Hogerheijde et al.\ 1999, Harvey et al.\ 2003, Looney, Mundy \& Welch 2003).

Harvey et al.\ (2003) recently analyzed observations of dust continuum 
emission at 1.2~mm and 3.0~mm from B335 on scales from 5000 to 500~AU.
A single radial power law in density provided a good description of the 
data, with best fit power law density index $p=1.65 \pm 0.05$. However,
this result is subject to the major systematic uncertainty of the unknown 
contribution from a central compact source, whose presence would bias 
the derived density profile to be too steep. Such a compact source might be 
expected in B335 if the protostar is surrounded by an accretion disk. 
Here, we present new IRAM PdBI observations that probe directly the 
subarcsecond millimeter continuum structure of B335. These observations
detect a compact source component and provide a measure of the previously 
uncertain level of point source ``contamination'' to the envelope emission, 
thereby allowing the inner density structure of B335 
to be constrained with unprecedented precision.

\section{Observations} 

Continuum emission from B335 was observed at 1.2~mm (246.5~GHz) and 
3.0~mm (100~GHz) with the six antenna IRAM PdBI in a modified 
A~configuration on 2003 March 05. Table~1 lists the observational parameters.
The visibilities span a baseline range from 23 to 400~meters, 
the longest baselines available. 
This range overlaps with the earlier D~configuration observations 
described by Harvey et al.\ (2003), thereby enabling cross calibration.
The pointing center and observing frequencies were chosen to be identical 
to the D~configuration observations, to facilitate the joint analysis 
of the two datasets. 
The half-power field of view for the PdBI is $20''$ (5000~AU) at 1.2~mm 
and $50''$ (12500~AU) at 3~mm. The absolute flux scale was set using
observations of the standard source MWC~349, assumed to be 1.77~Jy 
at 1.2~mm and 1.03~Jy at 3~mm. The estimated uncertainty in the 
flux scales is roughly 20\%. Frequent observations of nearby calibrators 
J1751+096 and J1925+211 were used to determine time-dependent complex gains.
Continuum visibility records were formed for each 60~s integration of the 
digital correlator, with 560~MHz bandwidth at 1.2~mm, and 570~MHz bandwidth 
at 3~mm. The bandpass was measured with observations of the strong source 
3C~273. The data were calibrated using the IRAM software package {\em CLIC} 
and comprise a total of 6000 records at each wavelength. 
In addition to amplitude and phase,
each record contains a variance measure, 
determined from the system temperatures and antenna gains.

\section{Results and Analysis}

We use the techniques developed by Harvey et al.\ (2003) to analyze the 
structure of B335 using the interferometer measurements directly
in the visibility domain, without producing images.  While this approach 
is computationally intensive, it recognizes the limitations of standard 
Fourier inversion and deconvolution techniques, and it avoids many 
difficulties that may arise from the synthesized beam characteristics.

\subsection{Presence of a Compact Component}

Figure~1 shows the visibility data at 1.2~mm ({\em diamonds}) and 
3.0~mm ({\em squares}), after binning logarithmically in $(u,v)$ distance. 
The 1.2~mm data show a visibility profile that flattens with increasing 
baseline lengths beyond $\sim60$~k$\lambda$. The start of this change in 
slope is also apparent in the 3.0~mm data. We interpret the flattening in 
the profiles as the separation between the extended envelope that dominates 
the flux at short baselines from a compact source component that persists to 
the longest baselines. 

To determine the central position, and to test whether the compact 
component is resolved, we fit an elliptical Gaussian to the long baseline 
visibilities, $ |(u,v)| > 100$~k$\lambda$. The fitted Gaussian is offset 
from the pointing center by 
($2\farcs27\pm0\farcs01, -0\farcs93\pm0\farcs02$), giving coordinates 
for the peak $\alpha = 19^{\rm h} 37^{\rm m} 00\fs89$, 
$\delta = 7^{\circ}34'10\farcs9$ (J2000)
consistent with earlier determinations of the protostar position.  
The uncertainty in the absolute position is likely considerably worse 
than the formal errors from this fit, since no extra effort was made 
to ensure especially accurate astrometry. The fitted Gaussian FWHM 
is $0\farcs36 \pm 0\farcs11 \: \times \: 0\farcs19 \pm 0\farcs06$, with 
position angle $14^{\circ} \pm 15^{\circ}$, which corresponds to 
semi-major axis $45 \pm 14$~AU and semi-minor axis $24\pm 8$~AU. 
The formal errors on the fitted size are at the $3\sigma$ level, 
and they are at the level of the seeing disk size $\sim0\farcs1$ 
estimated from Gaussian fits to the phase calibrators.

Figure~2 shows an image made from the 1.2~mm long baseline visibilities 
in the standard way, which highlights the compact source component.
We attribute essentially all of this compact flux component to thermal dust
emission. Reipurth et al.\ (2002) detected a subarcsecond scale radio 
source at 3.6~cm at this position, elongated in the outflow direction, with 
variable flux $<0.4$ mJy. But assuming the free-free emission from this 
ionized radio jet has a typical flat spectral index, it makes a negligble 
contribution to the emission at 1.2~mm. 
The shape of the sythesized beam matches the shape of the fitted Gaussian, 
but since the fit is to the raw visibility data, the sythesized beam does 
not directly affect the results, aside from there being inherently more 
resolution (longer baselines) East-West than North-South. The orientation of
the fitted Gaussian is consistent with a source that is elongated 
perpendicular to the axis of the bipolar outflow seen on larger scales 
(essentially East-West, e.g.\ Hirano et al.\ 1992). We therefore identify 
the compact component as a circumstellar 
disk. We caution that the fitted compact source size is only marginally 
resolved, and that the envelope emission may yet contaminate the signal, 
even on these subarcsecond size scales. In addition, if the compact 
component is a disk, then its outer radius is likely larger than the fitted 
Gaussian FWHM size, perhaps by a factor of up to $\sim 2$, because the 
emission profile of a disk with realistic density and temperature 
distributions is more centrally peaked than a Gaussian (Mundy et al.\ 1996, 
Wilner et al. 1996). This suggests a conservative upper limit of 100~AU for 
the disk radius.

\subsection{Calculation of Model Visibilities}

We compare the B335 visibilities from both antenna configurations 
to theoretical models of protostellar structure by constructing 
synthetic visibilities, taking account of 
(1) the dust continuum radiative transfer, and (2) the specifics of 
the observations, including the exact $(u,v)$ sampling and primary beam
attenuation. In brief, $1024 \times 1024$ pixel model intensity images 
(resolution $0\farcs1$~pixel${}^{-1}$ at 1.2~mm, $0\farcs2$~pixel${}^{-1}$ 
at 3~mm) are calculated using the full Planck function for the emissivity 
and integrating the radiative transfer equation through the model globule. 
Each model is normalized to a flux at 1.3~mm of $570 \pm 90$~mJy within a 
circular aperture (top-hat) of radius $20''$ based on the measurement of
Shirley et al.\ (2000). This normalization effectively provides a 
measurement at zero-spacing of the interferometer and is essential for
determining the envelope structure. Without this information, the model 
visibility profiles would float freely in normalization and the constraints 
that can be placed on the density structure are therefore much weaker 
(see e.g.\ Looney et al.\ (2003) where models with $p=2$ and $p=1.5$ 
generally cannot be distinguished.)
Observations are simulated by performing an FFT, assuming a Gaussian shape
for the primary beams (1.2~mm FWHM $20''$, 3~mm FWHM $50''$). 
The $(u,v)$ sampling is achieved by interpolating the real and imaginary 
parts of the resulting visibility grid. The center is fixed to match the 
position of the compact component from the elliptical Gaussian fit to the 
long baseline 1.2~mm data. 

{\em Compact Component.---} The compact component of dust emission in 
B335 evident at long baselines is likely due to a circumstellar disk. 
Since the size of the compact component from the Gaussian fit 
($\lesssim 0\farcs36$) is smaller than the fringe spacing on the 
longest baseline ($\sim 0\farcs6$), we calculate the contribution of this
compact component to the visibilities by modeling it as an unresolved 
point-like source. We assume the compact component to contribute a flux
$F_{\nu} \propto \nu^3$, for a disk that is optically thin at millimeter
wavelengths with a dust opacity index $\beta$ of unity 
(e.g.\ Beckwith et al.\ 1990).
If instead the disk were partly optically thick at these wavelengths, 
then its spectral index might be closer to 2. Experiments with the 
fitting procedure showed that both the inferred 1.2~mm flux of the compact 
component and the envelope density distribution are not sensitive to 
this spectral index assumption, and we adopt the optically thin model.
This lack of dependence on the disk spectral index occurs because the flux
of the compact component at 3~mm is not tightly constrained due to the lower
spatial resolution at this longer wavelength (see Figure~1).

{\em Envelope Density Models.---} 
We fix attention on the spherically symmetric broken power law models 
for the protostellar envelope density distribution, based on the success 
of this description in matching the previous lower resolution data.
The effect on the inferred density profiles of departures from spherical 
symmetry caused by the bipolar outflow was investigated by Harvey et al.\ 
(2003) and found to be small compared to other sources of systematic 
uncertainty (e.g.\ the dust temperature distribution).
The density models are of the form:
\begin{equation}
\rho_{d}(r) = \rho (R_0)
\begin{cases}
(r/R_0)^{-p} &  \mathrm{for \ } r \le R_0\\
(r/R_0)^{-2} &  \mathrm{for \ } R_0 < r \le R_{\mathrm{out}}
\end{cases}
\end{equation}
The fixed outer power law index of $2.0$, turn-over radius of $R_0=6500$~AU 
($26''$), and outer radius of $R_{\mathrm{out}}=0.15$~pc are based on 
the near-infrared dust extinction measurements of Harvey et al.\ (2001). 
The results are not sensitive to the outer boundary assumption due to 
the small field of view of the PdBI antennas. 

{\em Envelope Dust Temperature.---} The models necessarily include 
assumptions about the dust temperature distribution,
in addition to the density field. Based on the study of 
Shirley et al.\ (2002), we adopt the following for the temperature: 
\begin{equation}
T_{d} (r) = 
\begin{cases}
10 \; (5000 \; \mathrm{AU}/r)^{0.4} \; \mathrm{K} \ \ & \mathrm{for \ r}  
\le R_T\\
10 \; \mathrm{K} & \mathrm{for \ r} > R_T
\end{cases}
\end{equation}
i.e\ the temperature falls as a power law, reaching a constant value of 
$T=10$~K for radii beyond $R_T=5000$~AU. This is a close approximation to 
the results of detailed calculations of balanced heating and cooling.
For B335, an inner breakdown to this behavior is expected within a radius 
of $\lesssim 100$~AU, where the temperature gradient rises sharply due to 
the envelope becoming optically thick to infrared radiation. This scale is
at the resolution limit of the new data, so emission from this warm central 
region would appear as an unresolved point source in our visibility data.
Since the compact component of the flux is constrained by the long baseline 
data, this possible breakdown in the temperature distribution does not 
affect the inferred density profile of the envelope (for $r > 100$~AU), 
although it may contaminate the flux that we attribute to a circumstellar 
disk. To investigate the degree of envelope contamination, we assume that 
the envelope density distribution extends to the
center, and that the temperature increases steeply within a radius of 60~AU
(Shirley et al.\ 2002). In this case the mass of the warm region is only 
$5\times10^{-4}$~M$_{\odot}$. To contaminate the compact flux by 50\% 
requires the temperature to increase within this region by $\sim 230$~K 
above the assumed distribution, equivalent to an average temperature in the
region of $\gtrsim 300$~K. If the envelope density distribution
is disrupted so that it does not extend to the protostar location, then
it is even harder for the envelope to contaminate the disk flux.

{\em Envelope Dust Opacity.---} We assume that the dust opacity does not 
change with radius, although this approximation will likely break down in 
the innermost regions of the envelope where the temperature becomes high
enough that the ice mantles evaporate (Ossenkopf \& Henning 1994). 
For the dust opacity at millimeter wavelengths, we assume a power law 
$\kappa_{\nu} \propto \nu^{\beta}$ with index $\beta=1$ (e.g.\ Looney, 
Mundy, \& Welch 2000).  In the absence of absolute calibration 
uncertainties, the spectral index $\beta$ would control the relative 
normalization of the 1.2~mm and 3~mm profiles. In practice, the relative 
normalization of the 1.2~mm and 3~mm profiles from each of the antenna 
configuration are consistent with the assumed index, taking account the 
systematic uncertainties in the calibrations.

\subsection{Fitting Model Parameters and Evaluating Fit Quality}

We analyze the individual visibility records, and avoid binning the data 
to prevent loss of information. We maximize the probability distribution 
of a simultaneous fit to both datasets, by minimizing a modified $\chi^2$ 
of the form:
\begin{equation}
\tilde{\chi}^2 = \sum_{\lambda}{ \sum_{k=A,D}{ \left[
	\frac{(M_{\lambda,k}-1)^2}{\sigma(M_{\lambda,k})^2} + \sum_i{ 
	\frac{|Z_i-M_{\lambda,k} Z_i^{\mathrm{mod}}(p)|^2}{\sigma_i^2} } 
	\right] }}
\end{equation}
where the $Z_i$ are the visibility data points with uncertainty $\sigma_i$,
and the $Z_i^{\mathrm{mod}}(p)$ are the data points for a model with power 
law index $p$. The sum extends over both wavelengths, and over observations 
from both array configurations. The parameters $M_{\lambda,k}$ allow us to 
include the observational uncertainties, namely the uncertainty in the flux 
calibration of each dataset (each wavelength from each configuration), and 
the $\sim 15\%$ uncertainty in the Shirley et al.\ (2000) measurement used 
in the flux normalization. This is achieved by allowing the model 
visibilities to be scaled by the constrained parameter $M_{\lambda,k}$, 
assumed to be a Gaussian random variable with mean $1.0$ and standard 
deviation $\sigma(M_{\lambda,k})$. We adopt 
$\sigma(M_{1.2\mathrm{mm},k})=0.25$ for the two 1.2~mm datasets, making the 
assumption of two independent Gaussian distributions contributing to the 
probability distribution of each scaling factor. While this is not quite
accurate (since the normalization is part of both $M_{1.2\mathrm{mm},A}$ and
$M_{1.2\mathrm{mm},D}$), this formulation reduces the number of fitted
parameters and provides for realistic values of the scaling factors. 
These scaling factors essentially provide for the best possible 
cross-calibration for the observations from the two antenna configurations.
For the two 3~mm datasets, we allow the scaling parameters to be free, 
by adopting $\sigma(M_{3\mathrm{mm},k}) \rightarrow \infty$, essentially 
to account for the uncertain dust opacity spectral index. 

Since the models are non-linear in the fitting parameters, we analyze the 
uncertainty in the best-fit model parameters using the Monte Carlo technique
known as the {\em bootstrap} (Press et al.\ 1992). In brief, the dataset is 
resampled $n$ times, each time the fitting process is repeated and the 
best-fit parameters recorded, until the distribution of best-fit parameters 
is well sampled. The width of the distribution provides an estimate of the 
uncertainty in the parameters that best fit the original dataset.

The best fit model has envelope density power law index $p=1.55 \pm 0.04$ 
and a point source component of flux $F_{1.2\mathrm{mm}}=21 \pm 2$~mJy.
The constrained scaling parameters at 1.2~mm are 
$M_{1.2\mathrm{mm},A}=1.4 \pm 0.1$ and
$M_{1.2\mathrm{mm},D}=0.77 \pm 0.06$. 
The free scaling parameters at 3~mm are 
$M_{3\mathrm{mm},A}=1.4 \pm 0.2$ and  
$M_{3\mathrm{mm},D}=1.09 \pm 0.15$.
The unscaled best-fit model is offset to higher normalization than the 
D~configuration 1.2~mm data, and to lower normalization than the 
A~configuration 1.2~mm data. We interpret these results as most likely 
due to the uncertainties in the overall flux calibration of the two 
datasets, low for the former and high for the latter. The magnitudes of the 
scaling factors are consistent with this interpretation. 
In this context, the free scalings for the 3~mm datasets are close enough 
to unity that the data are consistent with an assumed dust opacity spectral 
index of unity. Figure~1 shows a plot of the various visibility datasets, 
renormalized by $1/M_{\lambda,k}$, binned in $(u,v)$ distance, along with 
the best fit model, and the two components of the best fit model. 

\section{Discussion}

\subsection{Implications for B335 Envelope Structure}

We find that the density profile of B335 has power law index 
$p=1.55 \pm 0.04$ within the inner $\sim 5000$~AU. This result is remarkably 
consistent with the $p=1.50$ expectation for free-fall collapse. The overall 
density structure
of B335 can be obtained by combining the inner behavior determined in the
present study with the steeper outer falloff determined from near-infrared
extinction work (Harvey et al.\ 2001) and low resolution millimeter and
submillimeter studies (Shirley et al.\ 2000, Motte \& Andr\'{e} 2001).
The extinction data showed an 
average power law index of $p=1.91 \pm 0.07$ over the region 3500-25,000~AU,
consistent with the $p=2$ index characteristic of a hydrostatic region
supported by thermal pressure (the index is not significantly affected 
by the small overlap with the region in which the index decreases). 
Joining these two regions together, for the adopted temperature distribution 
and mass opacity at 1.2~mm, the density distribution of molecular hydrogen 
is:
\begin{equation}
n_{H_2}(r) \simeq 3.3 \times 10^{4} \: \mathrm{cm}^{-3}
\begin{cases}
\left( \frac{r}{6500 \: \mathrm{AU}} 
\right)^{-1.5} & \ \ 100  \mathrm{~AU} \leq r \leq 6500   \mathrm{~AU}\\
\left( \frac{r}{6500 \: \mathrm{AU}} 
\right)^{-2.0} & \ \ 6500 \mathrm{~AU} <    r \leq 25,000 \mathrm{~AU}
\end{cases}
\end{equation}
The corresponding enclosed mass distribution is:
\begin{equation}
M(r) \simeq
\begin{cases}
0.58 {\rm M}_{\odot} \left( \frac{r}{6500 \: \mathrm{AU}} \right)^{1.5} & 
	\ \ 100  \mathrm{~AU} \leq r \leq 6500   \mathrm{~AU}\\
0.58 {\rm M}_{\odot} + 0.87 {\rm M}_{\odot} 
\left( \frac{r}{6500 \: \mathrm{AU}} - 1 \right) & 
	\ \ 6500 \mathrm{~AU} <    r \leq 25,000 \mathrm{~AU}
\end{cases}
\end{equation}
If one accounts for the asymmetry of the B335 core by modeling the outflow 
with a bipolar hollow cone, then the normalization of the density 
distribution away from the outflow must be increased by a small factor, 
$\sim 14$\% for a $40^{\circ}$ semi-opening angle.

This determination of the inner density structure of B335 has a precision 
that is unique amongst protostellar cores, and it provides the best 
evidence yet for the standard picture of isolated star formation, whereby
the inner regions of an initially hydrostatic isothermal envelope collapse
onto the center in near free-fall conditions (Shu, Adams \& Lizano 1987). 
%The density and implied velocity structure are consistent with the results 
%of applying the Shu (1977) inside-out collapse model to 
%molecular line observations (Zhou et al.\ 1993, Choi et al.\ 1995).
As noted by Harvey et al.\ (2003), while the B335 density distribution is 
qualitatively similar to the density distribution of the inside-out collapse 
model, the two distributions are quantitatively different in the 
inner regions. In the inside-out collapse model, the local density gradient 
within the infall radius is significantly less steep than $p=1.5$, 
ranging from roughly unity to 1.5. Such a shallow density profile is 
ruled out by the dust emission analysis.

The new long baselines observations reveal the contribution of the 
central compact source to the total dust emission and thereby remove the 
previously dominant source of systematic error from the envelope density 
structure determination (see Harvey et al.\ 2003). 
The remaining sources of systematic error, of which the most important 
is the detailed shape of the temperature distribution, likely contribute 
a systematic uncertainty of $\delta p \lesssim 0.15$.
In the future, high resolution observations of B335 at submillimeter 
wavelengths may help to mitigate the uncertainties associated with the 
temperature distribution. 
Spatially resolved information on the optically thin emission distribution 
at a third wavelength that is sufficiently short to be outside the 
Rayleigh-Jeans regime for the majority of the dense core should allow for 
partly breaking the degeneracy between temperature, density and opacity. 
Such observations may be obtained with the Submillimeter Array (SMA) on 
Mauna Kea, which is soon to be commissioned, and eventually the 
Atacama Large Millimeter Array (ALMA) in Chile, now under construction.
It is also clearly desirable to make high resolution spectral line 
observations of species not subject to depletion and unaffected by 
the bipolar outflow in order to best probe the dense core dynamics.

\subsection{Disk Properties}

The fitted flux of $F_{1.2\mathrm{mm}}=21 \pm 2$~mJy from the B335 disk 
is typical of the (distance corrected) $\sim 5$ to 300~mJy range of fluxes 
from disks around the T~Tauri stars observed by Beckwith et al.\ (1990)
and Osterloh \& Beckwith (1995). The disk flux is at the low end of the 
range for the disks in the Class~0 sources observed in the Perseus region 
by Brown et al.\ (2000) using the CSO-JCMT interferometer at 0.87~mm and by 
Looney, Mundy, \& Welch (2003) using BIMA at 2.7~mm (for $\beta\approx 1$).
However, the Perseus sample is highly biased, as the sources were chosen 
on the basis of their large millimeter fluxes. 

The $\sim 45$~AU FWHM spatial scale of the compact component in B335 at 
1.2~mm is similar to the sizes measured for the Perseus Class~0 disks 
(Brown et al.\ 2000), though it is not clear how accurately the Gaussian 
size measurement predicts the actual B335 disk radius. Given that disks
result from conservation of angular momentum during the collapse process, a 
small disk radius and disk flux might be expected in B335 because of its 
relatively low rotation rate (Frerking, Langer \& Wilson 1987, Zhou 1995). 
In the context of the Terebey, Shu \& Cassen (1984) model of inside-out 
collapse with rotation, the centrifugal radius for B335 is predicted to be 
only 3~AU, or $0\farcs012$, far too small to resolve. However, such a small 
disk could not produce the observed 1.2~mm flux without an unfeasibly high
dust temperature of $\gtrsim 1000$~K. The radius of the B335 disk is clearly
much larger than predicted by the simple model of inside-out collapse with
rotation.

The flux from the disk provides a measure of the disk mass, though the
large uncertainty in the normalization of the dust opacity dominates the 
mass calculation. Assuming $\kappa_{1.2~\mathrm{mm}}=0.02$~cm$^2$~g$^{-1}$, 
a simple estimate is 
$M_{\mathrm{disk}} = 0.004\, \mathrm{M}_{\odot} 
(50~\mathrm{K} / \langle T_{\mathrm{disk}} \rangle)$, 
where $\langle T_{\mathrm{disk}} \rangle$ is the mass weighted 
mean temperature. Theoretical expectation for a flared, 
radiatively heated disk, suggest 
$\langle T_{\mathrm{disk}} \rangle \simeq 50$~K (Beckwith 1999), though 
fits to the spectra of the T~Tauri disks in Beckwith et al.\ (1990) 
give temperatures that are generally lower by a factor of 2 to 3, and disk 
mass estimates that are higher by the same factor (occupying a range 
$\sim 0.002$ to 0.3~M$_{\odot}$). 
For B335, the appropriate mean temperature may be higher than for a disk 
around a T-Tauri star because of 
the surrounding envelope with substantial optical depth in the infrared 
that inhibits cooling (Natta 1993). 
But since the B335 disk flux is typical of T~Tauri stars, 
it seems likely that the disk mass is not unusually small or large.
Given the link between accretion disks and bipolar outflows, perhaps this is 
not surprising given that B335 exhibits a bipolar outflow that is not at all 
unusual, with size, momentum and energy typical for a low mass protostar 
(Goldsmith et al.\ 1984).

\section{Summary}

We present new subarcsecond resolution dust continuum observations of 
the protostellar collapse candidate B335 made with the IRAM PdBI 
at wavelengths of 1.2 and 3.0~mm. We analyze these data together 
with previous PdBI observations reported by Harvey et al.\ (2003) 
that provide short-baseline information. In summary:

\begin{enumerate}

\item The PdBI visibility datasets span baseline ranges from 15 to 400 
meters, and the longest baselines marginally resolve a compact component 
distinct from the highly resolved protostellar envelope. We identify
this compact component with a circumstellar disk, although emission from a
warm inner region of the envelope may also contribute to this flux. We 
place a conservative upper limit on the disk radius of $\simeq 100$~AU, 
given the uncertainties in the details of its temperature and density 
structure.

\item We analyze the visibility data by comparison with synthetic 
observations constructed from models with a variety of physical conditions.
We simultaneously constrain the envelope density distribution at radii 
from $\sim 100$ to $\sim 5000$~AU together with a point-like component 
to account for the disk.  The best fit model has an envelope density 
power law index $p=1.55 \pm 0.04$ within this region, with point source flux 
$F_{1.2\mathrm{mm}}=21 \pm 2$~mJy (1~$\sigma$).
Systematic uncertainties dominate the random uncertainties for the 
inferred density structure and contribute an estimated uncertainty of
$\delta p \lesssim 0.15$. 
The dominant source of systematic error is most likely the 
simple treatment of the dust temperature distribution.
Observations at submillimeter wavelengths may help to mitigate the 
uncertainties associated with the temperature distribution.

\item
The density structure of the B335 core indicates an $r^{-1.5}$ inner region 
in gravitational free-fall surrounded by an $r^{-2}$ envelope. 
However, the specific inside-out collapse solution of Shu (1977) does not
reproduce the data because of the shallow slope ($p \sim 1$) of the 
inside-out collapse solution just within the infall radius.

\item The fitted characteristics of the compact source suggest the 
B335 disk is typical of disks observed around T~Tauri stars, with 
radius $r \lesssim 100$~AU and mass 
$M \simeq 0.004$~M$_{\odot} \, (50$~K$/T_{\mathrm{disk}}$).
\end{enumerate}

\acknowledgements
We thank Diego Mardones for his efforts on the D-configuration dataset.
We acknowledge the IRAM staff from the Plateau de Bure and Grenoble for
carrying out all the observations and for their help during the data 
reduction. Partial support for this work was provided by NASA Origins of 
Solar Systems Program Grant NAG5-11777,

\clearpage
\begin{deluxetable}{lcc}
\tablenum{1}
\tabletypesize{\footnotesize}
\tablewidth{0pt}
\tablecaption{Summary of instrumental parameters}
\tablehead{\colhead{Parameter} & \colhead{1.3~mm} & 
	\colhead{3.0~mm}}
\startdata
Observation date & \multicolumn{2}{c}{2003 Mar.\ 05}\\
Configuration & \multicolumn{2}{c}{modified~A (six antennas)}\\ 
Baseline range & \multicolumn{2}{c}{23--400~m}\\
Pointing center (J2000) & 
 \multicolumn{2}{c}{$19^{h}37^{m}00\farcs74$, $+7^{\circ}34'10\farcs8$}\\
Phase calibrators & \multicolumn{2}{c}{J1751+096 \& J1925+211}\\
Bandpass calibrator & \multicolumn{2}{c}{3C~273}\\
Flux calibrator & \multicolumn{2}{c}{MWC~349}\\
Primary beam FWHM & $20''$ & $50''$\\
Observing frequency & 246.5~GHz & 100.0~GHz\\
%Bandwidth & 560~MHz & 570~MHz\\
RMS Noise & 1.9~mJy & 0.7~mJy\\
\enddata
\end{deluxetable}

\clearpage
%Figures
\begin{figure}
\figurenum{1}
\epsscale{0.95}
\plotone{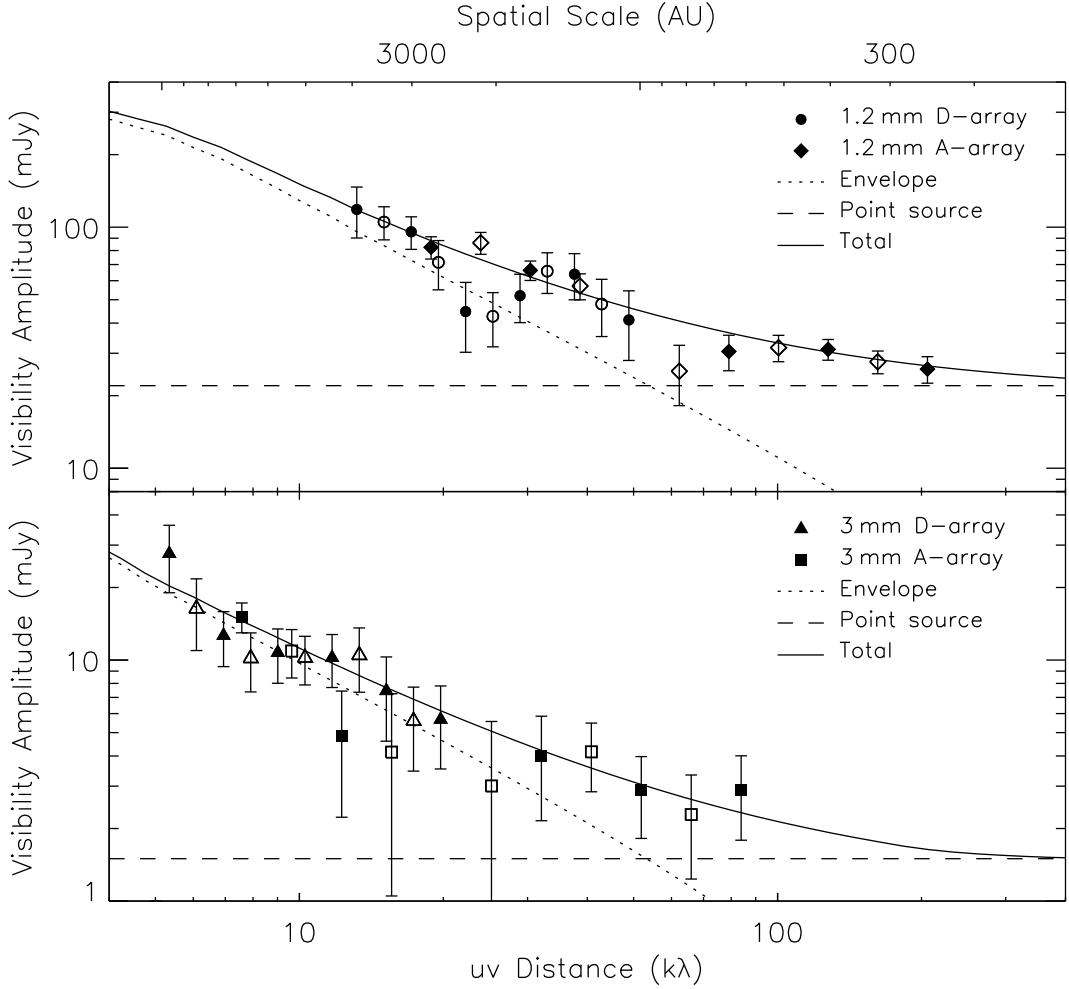}
\caption{Binned visibility amplitude vs.\ $(u,v)$ distance at 1.2~mm (upper 
panel) and 3.0~mm (lower panel) for the new IRAM PdBI observations of B335
together with the previous observations reported by Harvey et al.\ (2003).
The derived calibration scalings have been applied to the two datasets. 
Note that for each dataset the bins oversample the data, and therefore the 
filled symbols are not completely independent from the open symbols. 
The best-fit model (solid line) comprises a power law density distribution 
in the inner part of the envelope with $p=1.55 \pm 0.05$ (dotted line) and a 
point source of flux $F_{1.2\mathrm{mm}}=21\pm2$~mJy at 1.2~mm (dashed 
line).}
\end{figure}

\clearpage
\begin{figure}
\figurenum{2}
\epsscale{0.95}
\rotatebox{-90}{
\plotone{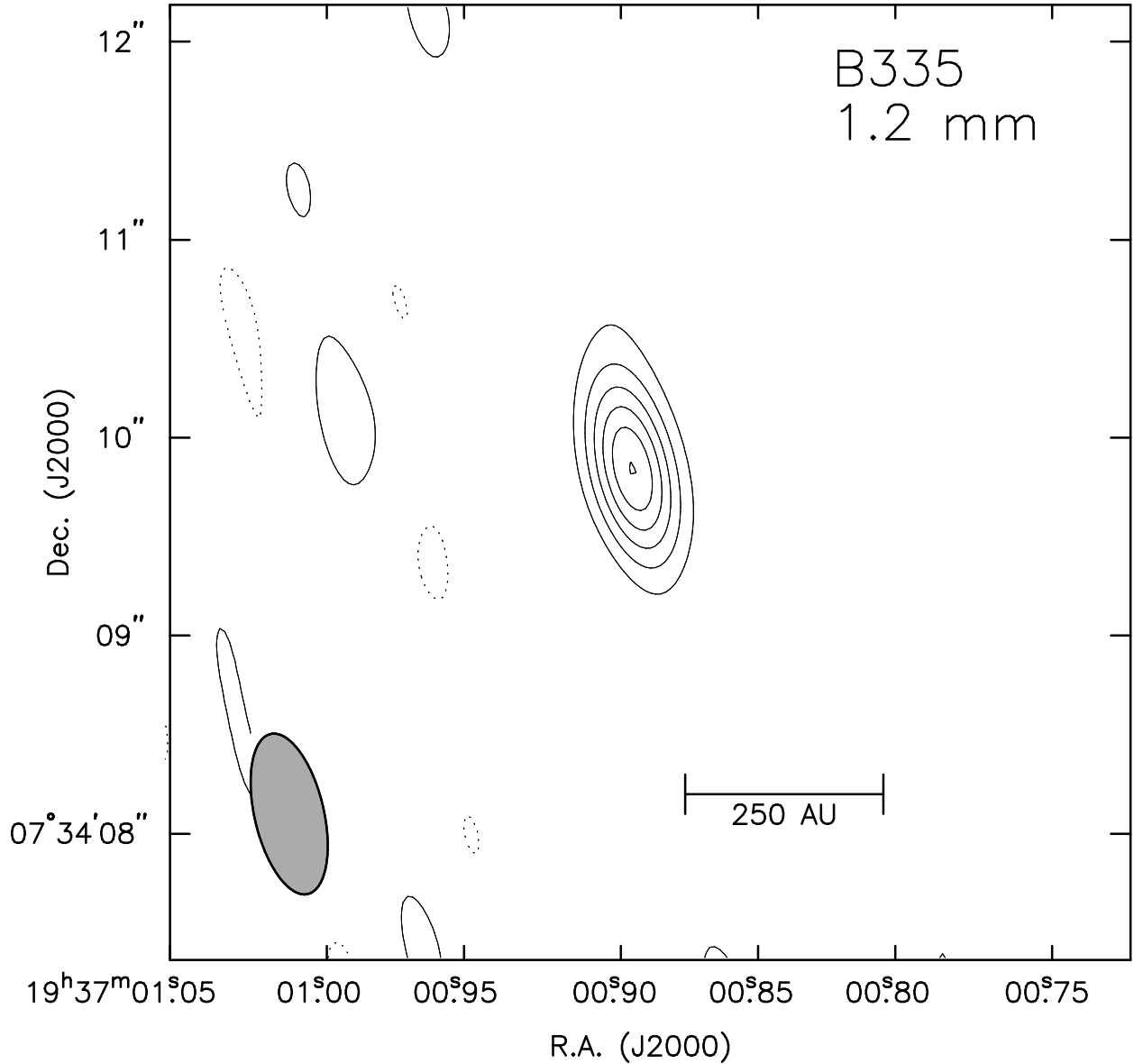}}
\caption{B335 image at 1.2~mm made from long baseline visibilities, 
$ |(u,v)| > 100$~k$\lambda$. 
The contour levels are $\pm2,4,6,...\times2.1$~mJy.
Negative contours are dotted. The ellipse in the lower left corner shows the
$0\farcs83\times0\farcs35$ p.a.\ $13^{\circ}$ synthesized beam. There is 
significant flux in a nearly pointlike component with maximum dimension 
similar to that of known circumstellar disks.}
\end{figure}

\end{document}